\documentstyle[12pt]{article}
\def\newline{\relax\ifhmode\null\hfil\break\else\nonhmodeerr@\newline\fi}
\def\frac#1#2{{#1\over#2}}
\def\text#1{{\hbox{\rm #1}}}
\def\flushpar{{\par \noindent}}

\newcommand{\beq}{\begin{equation}}
\newcommand{\eeq}{\end{equation}}
\newcommand{\bea}{\begin{eqnarray}}
\newcommand{\eea}{\end{eqnarray}}
\def\Id{ \mbox{1\hspace{-1.2mm}I} }
\def\nab#1{{\nabla_{#1}}}
\def\nabstar#1{\nabla\kern-0.5pt\smash{\raise 4.5pt\hbox{$\ast$}}
               \kern-4.5pt_{#1}}
\def\BE{\begin{equation}}
\def\EE{\end{equation}}
\def\BA{\begin{eqnarray}}
\def\EA{\end{eqnarray}}
\def\BAN{\begin{eqnarray*}}
\def\EAN{\end{eqnarray*}}
\def\LL{\left}
\def\RR{\right}
\def\nn{\nonumber\\}

\def\tr{\mbox{tr}}

\def\det{\mbox{det}}

\def\gm5{\gamma^5}

\def\bpsi{\bar{\psi}}

\def\CT{{\cal T}}
\def\anx{{\cal A}_n(x)}
\def\ancx{{\cal A}_c(x)}
\input epsf.sty
\newdimen\psfigsize
\def\psfigure#1 #2 #3 #4 #5{
    \begin{figure}[tbh]
      \begin{center}
      \vbox{
        \null\vskip-0.2in\hskip#2
        \epsfxsize=#1
        \epsfbox{#4}
        \vskip -0.3in
        \caption {#5 \label{#3}}
        \vskip 0.0 true in plus 0.3 true in
      }
      \end{center}
   \end{figure}
}
\begin{document}
\thispagestyle{empty}
\begin{flushright}
NTUTH-99-033 \\
June 1999
\end{flushright}
\bigskip\bigskip\bigskip
\vskip 2.5truecm
\begin{center}
{\LARGE {Fermion determinant and chiral anomaly \\
          on a finite lattice  }}
\end{center}
\vskip 1.0truecm
\centerline{Ting-Wai Chiu}
\vskip5mm
\centerline{Department of Physics, National Taiwan University}
\centerline{Taipei, Taiwan 106, Republic of China.}
\centerline{\it E-mail : twchiu@phys.ntu.edu.tw}
\vskip 2cm
\bigskip \nopagebreak \begin{abstract}
\noindent

The fermion determinant and the chiral anomaly of lattice
Dirac operator $ D $ on a finite lattice are investigated.
The condition for $ D $ to reproduce correct chiral anomaly at each
site of a finite lattice for smooth background gauge fields
is that $ D $ possesses exact zero modes satisfying the Atiyah-Singer
index theorem. This is also the necessary condition for $ D $ to have
correct fermion determinant ( ratio ) which plays the important
role of incorporating dynamical fermions in the functional integral.
We outline a scheme for dynamical fermion simulation of
lattice QCD with topologically proper $ D $.


\end{abstract}
\vskip 1.5cm

\newpage\setcounter{page}1

\section{ Introduction }

The fermion determinant, $ \det(D) $, of a Dirac operator $ D $, is
proportional to the exponentiation of the one-loop effective action
which is the summation of any number of external sources interacting
with one internal fermion loop. It is one of the most crucial quantities
to be examined in any lattice fermion formulations. In numerical
simulations of lattice QCD, its ratio between two successive gauge
configurations determines whether the effects of dynamical quarks are
properly incorporated, which are believed to be cruical to bring the
{\em ab initio} predictions of lattice QCD to agree with the experimental
data. Another important quantity in QCD is the chiral anomaly which breaks
the global chiral symmetry of massless QCD through the internal
quark loop. In the case of the axial current coupling to two
external photons, the axial anomaly \cite{ABJ} provides a proper account
of the decay rate of $ \pi^{0} \to \gamma \gamma $.
For the flavor singlet axial current coupling to two gluons,
the presence of chiral anomaly but no associated Goldstone boson
posed the $ U_A(1) $ problem which was resolved by 't Hooft \cite{tHooft76},
after taking into account of the topologically nontrivial gauge
configurations, the instantons. This may also provide an explanation why the
$ \eta $ ( $ \eta' $ ) particle is much heavier than the $\pi$'s ( $K$'s  ).
Lattice QCD should be designed to provide nonperturbative and
quantitive answers to all these and related problems.
However, these goals could not be attained if the
lattice Dirac fermion operator does not reproduce correct chiral
anomaly or the fermion determinant ( ratio ) on a finite
lattice {\em without fine-tuning any parameters.}
Furthermore, lattice QCD should realize the spontaneous breaking of
chiral symmetry in massless QCD, which gives massless Goldstone bosons
and thus leads to a correct and detailed description of the properties
of the strong interactions at low energy \cite{weinberg96}.

A basic requirement for lattice Dirac operator $ D $ is that, on a
finite lattice with any prescribed\footnote{ An example of prescribed
background gauge field is given in the Appendix. }
smooth background gauge field which has a well-defined topological
charge, $ D $ possesses {\em exact}
zero modes satisfying the Atiyah-Singer index theorem ( i.e., $ D $ is
{\em topologically proper} ) without fine tuning any parameters.
Here we would like to emphasize the importance of having
exact zero modes on a finite lattice without fine tuning any parameters.
If the zero modes can only be obtained by fine tuning
parameters, then it is impossible to tune the parameters
for every gauge configuration generated during a Monte Carlo simulation.

For any $ D $ satisfying this requirement, the sum of the anomaly
function\footnote{ Here we have adopted the terminology used in
                   Chapter 22 of ref. \cite{weinberg96}.}
over all sites must be correct, and is equal to two times of the
topological charge of the background gauge field. If it turns out that the
anomaly function of $ D $ at some of the sites do not agree with the
Chern-Pontryagin density, we can perform the following {\em topologically
invariant} transformation\footnote{The form of Eq. (\ref{eq:twc}) is
similar to the general solution of Ginsparg-Wilson relation given in
ref. \cite{twc98:6a}. However, the transformation (\ref{eq:twc}) can
be used for {\em any} Dirac operators. }
\bea
\label{eq:twc}
\CT(R) :  \hspace{4mm} D'= \CT(R)[D] \equiv D ( \Id + R D )^{-1}
\eea
with some operator $ R $ such that $ D' $ is local, then the anomaly
function of $ D' $ would be in good agreement with the Chern-Pontryagin
density at each site\footnote{ Here we have assumed that the size of
the finite lattice is large enough such that the finite size effects
can be neglected, i.e., the size of the lattice is much larger than
the localization length of $ D' $. }. It suffices to choose $ R $ to be
a hermitian operator which is local in the
position space and trivial in the Dirac space. The terminology,
{\em topologically invariant}, refers to the property that {\em the zero
modes and the index of $ D $ are invariant under the transformation
(\ref{eq:twc})}. Note that it is not necessary to
perform any fine tunings of $ R $ since any local $ D' $ would give the
correct chiral anomaly. Evidently, the set of transformations,
$ \{ \CT(R) \} $, with the group multiplication defined by
\bea
\label{eq:mult}
( \CT(R_1) \circ \CT(R_2) ) [D] = \CT(R_1) [ \CT(R_2)[D] ],
\eea
form an abelian group with the group parameter space $\{ R \}$,
due to the following basic property,
\bea
\label{eq:abelian}
\CT(R_1) \circ \CT(R_2) = \CT( R_1 + R_2 ).
\eea
In the topologically trivial sector, $ D^{-1} $ exists and
(\ref{eq:twc}) becomes
\bea
\label{eq:DI}
D'^{-1} = D^{-1} + R.
\eea
Then there exists a unique
\bea
\label{eq:Rc}
R_c = - \frac{1}{2} \{ D^{-1}, \gamma_5 \} \gamma_5
\eea
such that
\bea
D_c^{-1} = D^{-1} + R_c
\eea
is chirally symmetric, i.e.,
\bea
\gamma_5 D_c^{-1} + D_c^{-1} \gamma_5 = 0.
\eea
From the physical point of view, we must require the existence of
a chirally symmetric $ D_c  = \CT(R_c) [D] $ such that in the free
fermion limit $ D_c(p) \to i \gamma_\mu p_\mu $ as $ p \to 0 $,
otherwise, $ D $ is irrelevant to massless QCD. In general, we assume
that there exists
\bea
\label{eq:DcRc}
D_c = D ( \Id + R_c D )^{-1}
\eea
such that
\bea
\label{eq:Dc}
D_c \gamma_5 + \gamma_5 D_c = 0.
\eea
If $ D ( D_c ) $ is topologically proper, then $ D ( D_c ) $ must
be free of species doubling, thus $ D_c $ is non-local, according to
the Nielson-Ninomiya no-go theorem \cite{no-go}.
Furthermore, $ D_c $ has singularities
in topologically nontrivial background gauge fields \cite{twc98:9a}.
Substituting (\ref{eq:DcRc}) into (\ref{eq:Dc}), we obtain
\bea
\label{eq:gwr}
D \gamma_5 + \gamma_5 D = - 2 D R_c \gamma_5 D.
\eea
This is the rejuvenated Ginsparg-Wilson relation \cite{gwr}.
In general, for {\em any} lattice Dirac operator $ D $, if the chirally
symmetric $ D_c = \CT(R_c)[D] $ exists, then $ D $ must satisfy the
Ginsparg-Wilson relation.
From this viewpoint, the Ginsparg-Wilson relation really does {\em not}
specify the important attribute of a Dirac operator, namely
its {\em topological characteristics} \cite{twc98:9a}. Only for $ D $
is {\em topologically proper}
( e.g., the overlap-Dirac operator \cite{hn97:7} ), $ D $ ( or its
transform $ D' = D ( \Id + R D )^{-1} $ ) can be used
for lattice QCD, then the attractive features pointed out in
refs. \cite{ph98:2,chand98:5} can be realized on a finite lattice.
If $ D $ is topologically proper but nonlocal,
the transformation (\ref{eq:twc}) can be used to obtain
a local $ D' $ such that the anomaly function at each site of
a finite lattice could agree with the corresponding Chern-Pontryagin
density for smooth background gauge fields. Moreover, the locality of
$ D' $ also implies that its fermion determinant ratio
( with the zero modes omitted ) would agree with the corresponding
value in continuum.

\section{ The Anomaly Function }

In general, we consider lattice Dirac operator $ D $ which breaks the
continuum chiral symmetry according to
\bea
\label{eq:DB}
D \gamma_5 + \gamma_5 D = B
\eea
where $ B $ is a generic operator which is usually taken to be irrelevant
( i.e., vanish in the limit $ a \to 0 $ ). However, we can construct the
chirally symmetric action
\bea
\label{eq:A_s}
{\cal A}_s = \bpsi D_s \psi
\eea
using the chirally symmetric part of $ D $,
\bea
\label{eq:Ds}
D_s =  \frac{1}{2} ( D - \gm5 D \gm5 ).
\eea
Here the Dirac, flavor and internal symmetry indices are all suppressed.
Then $ {\cal A}_s $ has the usual chiral symmetry, and the divergence
of the associated Noether current\footnote{Here we assume that it is
the flavor singlet of the flavor symmetry group.} is
\BA
\partial^{\mu} J^5_{\mu}(x) =
    \bpsi_x \gm5 ( D \psi )_x  + (\bpsi D)_x \gm5 \psi_x
     - \frac{1}{2} ( \bpsi B )_x  \psi_x
     - \frac{1}{2} \bpsi_x  ( B \psi )_x
\label{eq:div5_n}
\EA
which satisfies the conservation law,
\bea
\label{eq:Q5}
\sum_x \partial^{\mu} J^5_{\mu}(x) = 0.
\eea
As usual, the lattice is taken to be finite with periodic boundary
conditions, and $ \partial_\mu J_{\mu}^5 (x) $ is defined by the
backward difference of the axial current
\BA
\partial_\mu J_{\mu}^5(x)=
\sum_{\mu} \ [ J_{\mu}^5 (x) - J_{\mu}^5 ( x - \hat{\mu} ) \ ]
\label{eq:divJ5_def}
\EA
such that it is parity even under the parity transformation, and
the conservation law Eq. (\ref{eq:Q5}) is also satisfied.

If $ D $ {\em does not possess exact zero modes in the background gauge
field }, then the fermionic average of $ \partial^{\mu} J^5_{\mu}(x) $
can be evaluated as
\BA
\label{eq:div5_f}
\LL< \partial^{\mu} J^5_{\mu}(x) \RR> &=&
\frac{1}{Z} \int [d\psi][d\bpsi] \partial^{\mu} J^5_{\mu}(x)
\exp( - \bpsi D \psi )  \\
Z &=& \int [d\psi][d\bpsi] \exp( - \bpsi D \psi )
\EA
Using (\ref{eq:div5_n}), we obtain
\bea
\label{eq:divJ5_nz}
\LL< \partial^{\mu} J^5_{\mu}(x) \RR> =
 \frac{1}{2} \ \tr \left[ ( B D^{-1} )(x,x) + ( D^{-1} B )(x,x) \right]
\equiv {\cal A}_n (x)
\eea
where the trace runs over the Dirac, flavor and
internal symmetry indices. The RHS of Eq. (\ref{eq:divJ5_nz}) is identified
to be the anomaly function of $ D $. If the chirally symmetric
limit of (\ref{eq:twc}), i.e., $ D_c = \CT(R_c)[D] $ exists,
then $ D $ satisfies the Ginsparg-Wilson relation
(\ref{eq:gwr}) and the anomaly function becomes
\BA
\label{eq:ax_gw}
\anx = - \tr \left[ \gamma_5 ( R_c D ) (x,x)
                   +\gamma_5 ( D R_c ) (x,x)  \right]
\EA
However, for any $ D $ which does not possess exact zero modes
in the background gauge field, the sum of the anomaly function
(\ref{eq:divJ5_nz}) over all sites must vanish due to the conservation law,
Eq. (\ref{eq:Q5}),
\BA
\label{eq:A=0}
\sum_x \anx = 0 \ .
\EA
This implies that if $ \anx $ is not zero identically for all $ x $,
then it must fluctuate from positive to negative values with respect
to $ x $. The latter case is exactly what happens to the anomaly function
of the standard Wilson-Dirac fermion operator.

On the other hand, if $ D $ {\em possesses exact zero modes in
topologically nontrivial background gauge fields}, then $ D^{-1} $
is not well defined. In this case, one needs to introduce an infinitesimal
mass $ m $ and evaluate (\ref{eq:div5_f}) with $ D $ replaced by
$ \hat{D} = D + m f[D] $ ( where $ f[D] $ is any functional of $ D $,
which has eigenvalue one for the exact zero modes of $ D $ ),
and finally take the limit ( $ m \to 0 $ ), i.e.,
\BA
\label{eq:div5_fm}
\LL< \partial^{\mu} J^5_{\mu}(x) \RR> &=& \lim_{ m \to 0 }
\frac{1}{Z} \int [d\psi][d\bpsi] \partial^{\mu} J^5_{\mu}(x)
\exp( - \bpsi \hat{D} \psi )  \\
Z &=& \int [d\psi][d\bpsi] \exp( - \bpsi \hat{D} \psi )
\EA
Inserting (\ref{eq:div5_n}) into (\ref{eq:div5_fm}), we obtain
\BA
\LL< \partial^{\mu} J^5_{\mu}(x) \RR>
&=& \lim_{m \to 0 } \frac{1}{2} \ \tr \left[ ( B \hat{D}^{-1} ) (x,x)
                                  +( \hat{D}^{-1} B ) (x,x) \right] \nn
& & + 2 \sum_{s=1}^{N_{+}} [\phi_s^{+}(x)]^{\dagger} \phi_s^{+} (x)
    - 2 \sum_{t=1}^{N_{-}} [\phi_t^{-}(x)]^{\dagger} \phi_t^{-} (x)
\label{eq:divJ5_z}
\EA
where $ \phi_s^{+} $ and $ \phi_t^{-} $ are normalized eigenfunctions
of $ D $ with eigenvalues $ \lambda_s = \lambda_t = 0 $ and
chiralities $ +1 $ and $ -1 $ respectively.
The first term on the RHS of Eq. (\ref{eq:divJ5_z}) is
identified to be the anomaly function of $ D $,
\BA
\label{eq:ax_anomaly}
\anx = \lim_{m \to 0 } \frac{1}{2} \ \tr \left[ ( B \hat{D}^{-1} ) (x,x)
                                     +( \hat{D}^{-1} B ) (x,x) \right]
\EA
Then summing Eq. (\ref{eq:divJ5_z}) over all sites
and using Eq. (\ref{eq:Q5}), we obtain
\BA
\label{eq:index_thm}
N_{-} - N_{+} = \frac{1}{2} \sum_x  \anx
\EA
This is the index theorem for {\em any} lattice Dirac operator on a finite
lattice, in {\em any} background gauge field. In the case
$ D $ does not possess any exact zero modes, Eq. (\ref{eq:index_thm})
reduces to Eq. (\ref{eq:A=0}). Equation (\ref{eq:index_thm}) implies that
the sum of the anomaly function over all sites must be a well defined even
integer for any background gauge field, however, it does {\em not} necessarily
imply the Atiyah-Singer index theorem for smooth background gauge fields.
This can be seen as follows. Since $ D $ does not have exact zero modes
in a trivial gauge background, the index of $ D $ must be proportional
to the topological charge $ Q $ of the smooth background gauge field.
Now, if $ Q $ is an integer, then the proportional constant must be
an integer, otherwise their product in general cannot be an integer.
Denoting this integer multiplier by $ c[D] $, we have
\bea
\label{eq:cD}
N_{-} - N_{+} = c[D] \ Q \ n_f
\eea
where $ n_f $ is number of fermion flavors.
In particular, for $ n_f = Q = 1 $, $ c[D] = N_{-} - N_{+} $. Here we have
assumed that $ c[D] $ is constant for smooth background gauge fields.
This is a reasonable assumption since $ c[D] $ is an intrinsic
characteristics of $ D $. However, when the gauge field becomes rough,
we expect that Eq. (\ref{eq:cD}) would break down. If one insists that
Eq. (\ref{eq:cD}) holds even for {\em rough} gauge configurations,
then $ c[D] $ cannot be an integer constant due to the highly nonlinear
effects of the gauge field. It is easy to deduce that,
in general, $ c[D] $ is a {\em rational number} functional of $ D $, which
in turn depends on the gauge configuration, but it becomes an integer
constant only for smooth gauge configurations.
The {\em topological characteristics} of $ D $,
$ c[D] $, was first discussed in ref. \cite{twc98:9a}, and was
investigated in ref. \cite{twc98:10a} for the Neuberger-Dirac
operator. Equation (\ref{eq:cD}) constitutes the index theorem for
{\em any} lattice Dirac operator on a finite lattice, in {\em any}
background gauge field with integer topological charge.
We can classify $ D $ according to its response in a smooth nontrivial
background gauge field. If $ D $ does not possess any zero modes,
then $ c[D] = 0 $, $ D $ is called {\em topologically trivial}.
If $ c[D]=1 $, (\ref{eq:cD}) becomes Atiyah-Singer index theorem,
then $ D $ is called {\em topologically proper}.
If $ c[D] $ is not equal to zero or one, then $ D $ is called
{\em topologically improper}.

Now we try to obtain a general expression for the anomaly function
satisfying (\ref{eq:index_thm}) and (\ref{eq:cD}).
Consider the gauge configuration with constant field tensors,
e.g., $ F^{a}_{12}(x) =  \frac{ 2 \pi q_1 }{ L_1 L_2 } $,
$ F^{a}_{34}(x) =  \frac{ 2 \pi q_2 }{ L_3 L_4 } $ and other $F$'s are zero,
where $ q_1 $ and $ q_2 $ are integers. Then the topological charge of this
configuration is
\bea
\label{eq:Q_top}
Q = \frac{1}{32 \pi^2} \ \sum_x \epsilon_{\mu\nu\lambda\sigma} \
     F_{\mu\nu}^{a}(x) F_{\lambda\sigma}^{b}(x+\hat\mu+\hat\nu)
       \ \tr \{ t_a t_b \} = n \ q_1 \ q_2
\eea
which is an integer.
Here $ t_a $ and $ t_b $ are generators of the internal symmetry group
with the normalization $ \tr \{ t_a t_b \} =  n \ \delta_{ab} $.
If $ D $ is local, then it is reasonable to assume that $ \anx $ is
constant for all $ x $. From Eqs. (\ref{eq:index_thm})-(\ref{eq:Q_top}),
we obtain
\BA
\label{eq:anxc}
\anx = \frac{n_f}{16 \pi^2} \ c[D] \ \epsilon_{\mu\nu\lambda\sigma} \
         F_{\mu\nu}^{a}(x) F_{\lambda\sigma}^{b}(x+\hat\mu+\hat\nu)
          \ \tr \{ t^a t^b \} \ .
\EA
However, this implies that any local $ D $ would yield the correct
anomaly function on a finite lattice, up to the integer factor $ c[D] $.
In reality, this cannot be true in general. The crux of the problem is
due to the fact that when $ D $ is topological trivial, i.e. $ c[D]=0 $
in the index theorem (\ref{eq:cD}), the anomaly function $ \anx $ does not
necessarily vanish identically at each site, but it can fluctuate
from positive to negative values provided that its sum over all sites is
zero. This is the complementary solution to the homogeneous equation
$ \sum_x \anx = 0 $. Using Eq. (\ref{eq:divJ5_nz}), we can
write the general form of the complementary solution as
\BA
\label{eq:comp}
 {\cal A}^{(h)} (x) = \sum_\mu \ [ G_\mu(x) - G_\mu(x-\hat\mu) ] \equiv
\partial_\mu G_\mu (x)
\EA
where $ G_\mu(x) $ is any local and gauge invariant function.
Then the general solution of the anomaly function is the sum of the
complementary solution (\ref{eq:comp}) and the particular
solution (\ref{eq:anxc}),
\BA
\label{eq:anx}
\anx &=& \frac{n_f}{16 \pi^2} \ c[D] \ \epsilon_{\mu\nu\lambda\sigma} \
         F_{\mu\nu}^{a}(x) F_{\lambda\sigma}^{b}(x+\hat\mu+\hat\nu)
          \ \tr \{ t^a t^b \}   \nn
     & & + \sum_\mu \ [ G_\mu(x) - G_\mu(x-\hat\mu) ]
\EA
In general, $ G_\mu (x) $ depends on the background gauge field
as well as the lattice Dirac operator $ D $.
If $ {\cal A}^{(h)} (x) $ does not vanish in the presence of a constant
gauge background, then $ \anx $ must be different
from the continuum chiral anomaly, hence this $ D $ should be
abandoned. Only when $ {\cal A}^{(h)} (x) $ vanishes identically for any
smooth gauge background, $ D $ can reproduce
the correct chiral anomaly for $ c[D] = 1 $ ( e.g., the Neuberger-Dirac
operator ). On the other hand, the pathologies of the
standard Wilson-Dirac operator are : $ c[D] = 0 $ and
$ {\cal A}^{(h)} (x) \ne 0 $.

Now we consider $ D $ with anomaly function $ \anx $ satisfying
(\ref{eq:anxc}) in a constant gauge background.
Then we introduce local fluctuations [ e.g., the sinusoidal
terms in Eqs. (\ref{eq:A1})-(\ref{eq:A4}) ] to the background
gauge field and/or increase its topological charge $ Q $. Then $ c[D] $
in Eqs. (\ref{eq:anxc}) and (\ref{eq:cD}) may remain constant provided
that the local fluctuations of the gauge fields are not too violent and/or
$ Q $ is not too large. When the gauge background becomes so rough
that $ D $ is non-local, then Eq. (\ref{eq:anxc})
is no longer valid, but the index theorem, Eq. (\ref{eq:cD}) may still
hold with the same $ c[D] $. In this case, the anomaly function $ \anx $
can be expressed in terms of the general formula (\ref{eq:anx}) with the
same $ c[D] $ but nonvanishing $ \partial_\mu G_\mu (x) $.
If we keep on increasing the roughness of the gauge background,
then $ D $ will undergo a topological
phase transition, and $ c[D] $ in Eqs. (\ref{eq:cD}) and (\ref{eq:anx})
will become another integer or even a fraction ( Explicit examples of
topological phase transition will be demonstrated in Section 4 ).
In general, Eq. (\ref{eq:anx}) constitutes the anomaly function for
{\em any} lattice Dirac operator $ D $ on a finite lattice, in {\em any}
background gauge field. However, only when $ {\cal A}^{(h)} (x) = 0 $
and $ c[D] = 1 $, the anomaly function can agree with the
Chern-Pontryagin density.

If the chirally symmetric limit of $ D $,
i.e., $ D_c = \CT(R_c)[D] $, exists,
then $ D $ satisfies the Ginsparg-Wilson relation (\ref{eq:gwr}),
and its anomaly function (\ref{eq:ax_anomaly}) becomes exactly in
the same form of Eq. (\ref{eq:ax_gw}).
Substituting Eq. (\ref{eq:ax_gw}) into the LHS of Eq. (\ref{eq:anx}),
we obtain
\BA
\label{eq:anx_gw}
\anx &=& - \tr \left[   \gamma_5 ( R_c D ) (x,x)
                             + \gamma_5 ( D R_c ) (x,x) \right] \\
&=& \frac{n_f}{16 \pi^2} \ c[D] \ \epsilon_{\mu\nu\lambda\sigma} \
         F_{\mu\nu}^{a}(x) F_{\lambda\sigma}^{b}(x+\hat\mu+\hat\nu)
          \ \tr \{ t^a t^b \}                      \nn
& & + \sum_\mu \ [ G_\mu(x) - G_\mu(x-\hat\mu) ]
\label{eq:anomaly}
\EA
which agrees with the results obtained in ref. \cite{twc98:9a,twc99:1}.
For $ U(1) $ lattice gauge theory with "admissible" background
gauge configurations, Eq. (\ref{eq:anomaly})
agrees with the result derived by L\"uscher \cite{ml98:8b}.
For the Overlap-Dirac operator, in the weak coupling limit or the
classical continuum limit, we have $ \partial_\mu G_\mu (x) = 0 $,
then Eq. (\ref{eq:anomaly})
agrees with the results derived by Kikukawa and Yamada \cite{kiku98:6},
Fujikawa \cite{fuji98:11}, Adams \cite{adams98:12} and
Suzuki \cite{suzu98:12} respectively.

Inserting Eq. (\ref{eq:ax_anomaly}) [ Eq. (\ref{eq:ax_gw}) ] into
Eq. (\ref{eq:index_thm}), we obtain
\BA
\label{eq:index_thm_gw}
N_{-} - N_{+} = - \frac{1}{2} \sum_x
                  \tr \left[   \gamma_5 ( R_c D ) (x,x)
                             + \gamma_5 ( D R_c ) (x,x) \right] \
\EA
which agrees with the index theorem for Ginsparg-Wilson fermion
\cite{ph98:1,ml98:2}. Note that (\ref{eq:index_thm_gw})
does {\em not} necessarily imply the Atiyah-Singer index theorem
for smooth background gauge fields with integer topological charge.

If the lattice Dirac operator $ D $ does {\em not } have exact zero
modes in the presence of topologically non-trivial
background gauge fields, however, it manages to possess exact zero modes
in a certain limit, say, for example, the Wilson-Dirac fermion operator
\cite{wilson75} in the classical continuum limit ( i.e., zero lattice
spacing with infinite number of sites in each direction ).
Then on any finite lattice, this lattice Dirac operator cannot
reproduce the correct anomaly function for topologically non-trivial
background gauge fields, since the sum of the anomaly function
(\ref{eq:divJ5_nz}) over the entire lattice must be zero
due to the conservation law (\ref{eq:Q5}),
\bea
\label{eq:ax_zero}
\frac{1}{2} \sum_x \ \tr \left[ ( B D^{-1} )(x,x) + ( D^{-1} B )(x,x) \right]
= \sum_x \LL< \partial^{\mu} J^5_{\mu}(x) \RR> = 0
\eea
while the sum of Chern-Pontryagin density over all sites is proportional
to the topological charge,
\bea
\label{eq:rho_top}
\sum_x \frac{1}{16\pi^2} \epsilon_{\mu\nu\lambda\sigma} \
     F_{\mu\nu}^{a}(x) F_{\lambda\sigma}^{b}(x+\hat\mu+\hat\nu)
       \ \tr \{ t^a t^b \} = 2 \ Q
\eea
which is nonzero for topologically nontrivial background gauge fields.
As we increase the number of sites or decrease the lattice spacing,
the anomaly function does not approach the Chern-Pontryagin density
since it must satisfy the constraint Eq. (\ref{eq:ax_zero}).
It is evident that {\em the limit where $ D $ might possess exact
zero modes and correct anomaly function cannot be obtained by
smoothly extrapolating measurements at finite lattice spacings
towards $ a = 0 $ since the total anomaly (\ref{eq:ax_zero})
must undergo a discontinuous jump from zero to a nonzero integer
at the point $ a = 0 $, if the background gauge field is topologically
nontrivial}. Only for the trivial background gauge field,
$ D $ might have the chance to reproduce the correct anomaly by
extrapolations towards the limit $ a = 0 $.
If one attempts to bypass this difficulty by adding a mass term to $ D $
with the hope of obtaining the correct anomaly by tuning the mass and/or
other parameters, then one must encounter the unsurmountable problem of
fine-tuning, since it is almost impossible to tune the parameters
to some fixed values such that the Chern-Pontryagin density is
reproduced at each site, and for all different gauge configurations.
Therefore the chiral limit of these lattice Dirac operators
( e.g., the standard Wilson-Dirac fermion operator ) not only is
difficult to be realized in practice, but also in principle a
goal impossibly to be attained. On the other hand, if the lattice
Dirac operator ( e.g., the overlap-Dirac operator \cite{hn97:7} )
can reproduce exact zero modes satisfying the Atiyah-Singer index
theorem on a finite lattice, then the total anomaly
is bound to be correct, thus the anomaly function could agree
with the Chern-Pontryagin density even at finite lattice spacings.

\section {Fermion Determinant}

Now let us turn to the fermion determinant which plays the important
role of incorporating dynamical fermions in the functional
integral. Formally, the fermion determinant can be
represented as the product of all eigenvalues of $ D $.
In continuum, the fermion determinant must be regularized, and only the
ratio of fermion determinants can be well defined and independent of
regularization. In the topologically trivial sector, it is ususally
normalized to one in the free fermion limit, i.e.,
$ \det[D(A)] \leftarrow \det[ D(A) ]/ \det[ D(0)] $. However, in the
non-trivial sectors, $ D $ has exact zero modes, then the fermion
determinant ratio is well-defined only for gauge configurations within
the same topological sector or with the zero modes omitted.

On a lattice, if $ D $ is topologically proper, then the fermion
determinant in the functional integral also serves another important
role {\em to "decompose" configurations of link variables
into distinct topological sectors.}
This can be seen as follows.
On a lattice, any two gauge configurations are smoothly connected.
In Monte Carlo simulations, the transition probability between any two
gauge configurations in the quenched approximation is proportional to
the exponentiation of the difference of their actions, which is finite.
This is in contrast to the situation in the continuum, where the gauge
configurations are decomposed into disconnected topological sectors,
and the functional integral is performed in each topological sector
individually.
However, if $ D $ is topologically proper and its fermion determinant
is incorporated in the functional integral,
then the transition probability between any two gauge configurations
in two different topological sectors is either {\em zero} or {\em infinite},
in the limit $ m \to 0 $ ( Here $ m $ is an infinitesimal mass introduced
to $ D $ ). So, if one cutoffs all transitions having
a transition probability less than a certain lower bound, say, $ p_1(m) $, or
larger than a upper bound, say, $ p_2(m) $, then only gauge configurations
within the same topological sector can be accepted in the Monte
Carlo updatings, thus the functional integral can be evaluated exactly
in the same manner as in the continuum with disconnected topological
sectors. Therefore Monte Carlo simulations and measurements of
observables are performed separately for each relevant topological sector,
and the final result is averaged over all relevant topological sectors
with weight $ e^{i \nu \theta } $, where $ \theta $ denotes the vacuum angle
and $ \nu $ the winding number. From this viewpoint, the deficiency of link
variables being smoothly deformable to the identity element can be
circumvented by incorporating dynamical fermions in the functional
integral, provided that $ D $ is topologically proper.

To be specific, consider lattice QCD with $ n_f $ flavors of "massless"
quarks, where the lattice Dirac operator $ D $ is topologically proper.
Then the expectation value of a general observable
$ {\cal O}(U,\psi,\bar\psi) $ of the field varibales can be written as
\bea
\label{eq:OB}
\left< {\cal O} \right> =
\frac{1}{Z} \int [dU] [d\psi] [d\bar\psi] {\cal O}(U,\psi,\bar\psi)
\exp ( - {\cal A}_g - \bar\psi D(U) \psi)
\eea
where $ {\cal A}_g $ is Wilson's pure gauge action, and $ Z $ is
the partition function
\bea
\label{eq:Z}
Z = \int [dU] \det[ D(U) ] \exp ( - {\cal A}_g )
\eea
Here the functional integral of gauge fields must include the contributions
of instantons. In four dimensions, for any gauge group $ G $ which is
a simple Lie group ( e.g., $ G = SU(3) $ ), the instantons are characterized
by the elements of the homotopy group
$ \pi_3 (G) = \pi_3 (SU(2)) = {\cal Z } $,
i.e., the winding number,
\bea
\label{eq:winding}
\nu = \frac{1}{64\pi^2} \sum_x \epsilon_{\mu\nu\lambda\sigma}
         F^{a}_{\mu\nu}(x) F^{a}_{\lambda\sigma} (x+\hat\mu+\hat\nu)
\eea
which satisfies
\bea
\label{eq:index_nu}
n_f \ \nu = N_{-} - N_{+}
\eea
where $ N_{+} ( N_{-} ) $ denotes the number of zero modes of $ D $
of positive (negative) chirality.
Suppose that the weight factor for each topological sector is $ w(\nu) $,
then Eqs. (\ref{eq:OB}) and (\ref{eq:Z}) are replaced by
\bea
\label{eq:OBI}
\left< {\cal O} \right> &=&
\frac{1}{Z} \sum_{\nu} w(\nu)
\int [dU] [d\psi] [d\bar\psi] {\cal O}(U,\psi,\bar\psi)
\exp ( - {\cal A}_g - \bar\psi D(U) \psi) \\
Z &=& \sum_\nu w(\nu)
\int [dU] \det[D(U)] \exp ( - {\cal A}_g ) \ .
\label{eq:ZZ}
\eea
If we require $ \left< {\cal O } \right> $ to satisfy the
cluster decomposition principle \cite{weinberg96}, then it is
straightforward to deduce that
\bea
\label{eq:wnu}
w(\nu) = \exp( i \nu \theta )
\eea
where $ \theta $ is a free parameter,
the so called vacuum angle \cite{theta}.
Due to the presence of fermion determinant, the partition function
(\ref{eq:ZZ}) only receives the contribution from the trivial sector,
\bea
\label{eq:ZZZ}
Z = \int [dU]^{(0)} \det[D(U)] \exp ( - {\cal A}_g )
\eea
where the superscript $ (0) $ stands for the trivial sector.
Without loss of generality, we write the observable $ {\cal O} $ as
\bea
\label{eq:OU}
{\cal O }( U, \psi, \bar\psi ) =
{\cal O} (U) \psi_{x_1} \bar\psi_{y_1} \cdots  \psi_{x_n} \bar\psi_{y_n} \ .
\eea
Then, for the trivial sector, (\ref{eq:OBI}) can be evaluated as
\bea
\label{eq:OUT}
\left< {\cal O} \right> =
\frac{1}{Z}
\int [dU]^{(0)} \exp ( - {\cal A}_g ) \ {\cal O}(U) \ \det(D)
\sum_{z_1,\cdots,z_n} \epsilon^{z_1 \cdots z_n}_{x_1 \cdots x_n }
D^{-1}_{z_1 y_1} \cdots D^{-1}_{z_n y_n}
\eea
where $ \epsilon^{z_1 \cdots z_n}_{x_1 \cdots x_n } $ is the
completely antisymmetric tensor which is equal to $ +1 (-1) $ if
$ \{ z_1 \cdots z_n \} $ is even ( odd ) permutation of
$ \{ x_1 \cdots x_n \} $, and is zero otherwise.
For topologically non-trivial sectors, (\ref{eq:OBI}) can be
evaluated by expanding the fermion fields in terms of the complete
set of eigenfunctions of $ D $ with Grassman coefficients, then
the functional integral over the fermion fields can be expressed
in terms of that over the Grassman coefficients. It is evident that
for $ {\cal O} $ given in (\ref{eq:OU}), non-trivial sectors with
winding number $ | \nu | > n $ do not contribute to (\ref{eq:OBI}).
In general, (\ref{eq:OBI}) can be written as
\bea
\label{eq:OBF}
\left< {\cal O} \right> =
\frac{1}{Z} \sum_{\nu} \exp( i \nu \theta )
\int [dU]^{(\nu)} \exp ( - {\cal A}_g ) \
\det_1[D(U)] \ {\cal O}(U) \ {\cal F}(U)
\eea
where $ \det_1[D(U)] $ denotes the fermion determinant with the zero modes
omitted, and $ {\cal F} (U) $ the remnants of the observable involving
fermion fields after incorporating the zero modes from the fermion
determinant, which in general can be expressed in terms of the
eigenfunctions of $ D $. It is evident that the relevant topological
sectors which have nonzero contributions to (\ref{eq:OBF}) must
come in $ \pm \nu $ pairs ( $ \nu \ne 0 $ ).

Now consider Monte Carlo simulations of lattice QCD with dynamical
quarks. Suppose the initial gauge configuration is $ \{ U \} $
in the topological sector of winding number $ \nu $. A trial
configuration $ \{ U' \} $ with winding number $ \nu' $ is generated.
We intend to reject $ \{ U' \} $ if $ \nu' \ne \nu $.
Then, for $ \nu' = \nu $, $ \{ U' \} $ is accepted according to the
Metropolis algorithm. That is, we compute the ratio
\bea
\label{eq:prob}
    P( \{ U \} \rightarrow \{ U' \} )
& \equiv & \frac{ \det[ D(U')] }{ \det[ D(U)] }
           \exp( - {\cal A}_g (U') + {\cal A}_g (U) )
\eea
and compare $ P $ with a random number $ x $ drawn from the uniform
distribution in $ ( 0, 1 ) $. If $ x < P $, then $ \{ U' \} $ is
accepted, otherwise $ \{ U' \} $ is rejected. Now our problem is
how to reject $ \{ U' \} $ with $ \nu' \ne \nu $, before the
Metropolis sampling.

In general, for any two configurations $ \{ U \} $ and $ \{ U' \} $,
(\ref{eq:prob}) can be written as
\bea
\label{eq:probz}
  P( \{ U \} \rightarrow \{ U' \} )
= 0^{ n_f ( |\nu'| - |\nu| ) }
  \frac{ \det_1[ D(U')] }{ \det_1[ D(U)] }
  \exp( - {\cal A}_g (U') + {\cal A}_g (U) )
\eea
where $ \det_1 $ denotes the fermion determinant with zero modes
omitted\footnote{ Here we have assumed that $ D $ is normal
and thus $ \det_1(D) $ is positive definite.}.
It is evident that $ P $ is zero if $ |\nu'| > |\nu| $,
but $ \infty $ if $ |\nu'| < | \nu | $. So, we can easily reject
$ \{ U' \} $ in these two cases. Next we determine whether $ \nu' = \nu $
or $ \nu' = -\nu $. This can be done in the following.
In reality, the quarks may not be exactly massless.
However, if there exists a very light quark ( e.g. $ u $ or $ d $ quark )
of almost zero mass which enters the action in the form
$ m \bar\psi_R \psi_L + m^* \bar\psi_L \psi_R $ ( $ m^* $ is complex
conjugate of $ m $ ), then the factor
$ 0^{n_f(|\nu'|-|\nu|)} $ in Eq. (\ref{eq:probz}) is replaced by
\bea
\label{eq:mass}
R(m) \equiv \frac{ m^{\nu'} \theta(\nu') + {m^*}^{-\nu'} \theta(-\nu') }
                 { m^{\nu } \theta(\nu ) + {m^*}^{-\nu } \theta(-\nu ) }
\eea
where $ \theta(x) $ is the step function. In order to reject the
configurations $ \{ U' \} $ with winding number $ \nu' \ne \pm \nu $,
we first set $ m^* = m \sim 0 $, and compute $ P $ according to its
definition (\ref{eq:prob}).
Then we reject $ \{ U' \} $ if $ P < p_1(m) $ or $ P > p_2(m) $,
where $ p_1(m) $ and $ p_2(m) $ are easily determined since
$ P $ becomes either zero or infinity for $ \nu' \ne \pm \nu $,
in the limit $ m \to 0 $. Now we have $ \{ U' \} $ with $ \nu' = \pm \nu $.
Next we restore $ m^* \ne m $ and
compute $ P $ again using (\ref{eq:prob}). If $ P $ is real, then
$ \nu' = \nu $, according to (\ref{eq:mass}),
so $ \{ U' \} $ is further sampled by Metropolis algorithm.
On the other hand, if $ P $ is complex, then $ \nu' = - \nu $,
but the configuration $ \{ U' \} $ can be used for the sector of winding
number $ -\nu $ rather than being thrown away.
As mentiond above, in general, the relevant topological sectors for
observables in QCD must come in $ \pm \nu $ pairs. So, our strategy is
to perform Monte Carlo simulations for both $ \pm \nu $ sectors
simultaneously such that optimal efficiency can be attained.
After enough statistics have been accumulated for $ \pm \nu $ sectors,
then we can move on to another pair of relevant topological sectors by
starting with a prescribed gauge configuration.
Finally the expectation value $ \left< {\cal O} \right> $, (\ref{eq:OBF}),
can be obtained by averaging measurements over all relevant topological
sectors with weight factor $ \exp( i \nu \theta ) $.

On the other hand, if $ D $ does {\em not} possess exact zero modes, then
we cannot use the above prescription to decompose the link
configuartions into distinct topological sectors. Furthermore, if $ D $ is
topologically improper ( or trivial ), then its fermion determinant must be
very different from its counterpart in continuum. Even if we introduce a
mass term, the fermion determinant of these Dirac operators ( e.g., the
standard Wilson-Dirac fermion operator\footnote{ We recall that the
Wilson-Dirac fermion operator is topologically trivial,
which does not possess exact zero modes on a finite lattice.} )
does not tend to the correct value in the chiral limit,
thus the effects of dynamical fermion cannot be incorporated properly.
Therefore the {\em necessary} condition
for $ D $ ( or $ D'=D(\Id + R D )^{-1} $ ) to have
the correct fermion determinant is that $ D $ is topologically proper.
Then the locality of $ D' $ would in general lead to the correct
fermion determinant $ \det(D') $ on a finite lattice.
However, we still cannot formulate
a criterion ( such as the locality of a topologically proper $ D' $ )
that can gaurantee the correct fermion determinant on a finite lattice.
We are sure that, if such a criterion exists, it must be
more stringent than that for the chiral anomaly, since the
former involves all eigenvalues of $ D $ while the latter only the
zero modes.

A viable scheme to reproduce the continuum fermion
determinant on the lattice is Narayanan and Neuberger's
Overlap formalism \cite{rn95} which was inspired by
Kaplan's Domain-Wall fermion \cite{kaplan92} and
Frolov-Slavnov's generalized Pauli-Villars regularization \cite{slavnov93}
with an infinite number of Pauli-Villars fields.
The lattice Dirac operator proposed by Neuberger \cite{hn97:7} fulfils the
crucial requirements of reproducing correct chiral anomaly and
fermion determinant on a finite lattice, as reported in ref. \cite{twc98:4}.

\section{ Some Examples }

In this section, we illuminate above discussions by computing the
axial anomaly and the fermion determinant of a topologically proper
$ D $ which is constructed for the purpose of demonstration,
and compare with those of the standard Wilson-Dirac fermion operator
which is topologically trivial. For simplicity, we perform these
computations on two-dimensional lattices with $ U(1) $ background
gauge field, and compare the numerical results with the exact solution
on the torus. It is straightforward to extend these numerical
computations to four dimensional lattices with non-abelian background
gauge fields.

First, we set up the notations for the Wilson-Dirac fermion operator
$ D_w $ with Wilson parameter $ r_w > 0 $,
\bea
\label{eq:Dw}
D_w = m_f
      + \frac{1}{2} \left[ \gamma_{\mu} ( \nabstar{\mu} + \nab{\mu} ) -
                      r_w  \nabstar{\mu} \nab{\mu} \right]
\eea
where $ m_f $ is the fermion mass,
$ \nab{\mu} $ and $ \nabstar{\mu} $ are the forward and backward difference
operators defined as follows.
\bea
\nab{\mu}\psi(x) &=& \
   U_\mu(x)\psi(x+\hat{\mu})-\psi(x)  \nonumber \\
\nabstar{\mu} \psi(x) &=& \ \psi(x) -
   U_\mu^{\dagger}(x-\hat{\mu}) \psi(x-\hat{\mu})  \nonumber
\eea
From Eq. (\ref{eq:divJ5_nz}), the divergence of axial current for
the Wilson-Dirac fermion operator is
\bea
\label{eq:divJ5_dw}
\LL< \partial^{\mu} J^5_{\mu}(x) \RR> &=&
   2 m_f \ \tr \left[ \gm5 D_w^{-1} (x,x) \right] \nn
& & - \frac{r_w}{2} \ \tr \LL\{ \gm5 [ ( \nabstar{\mu} \nab{\mu} ) D_w^{-1}
                        + D_w^{-1} (\nabstar{\mu} \nab{\mu} ) ] (x,x) \RR\}
\eea
The last term in (\ref{eq:divJ5_dw}) is the axial anomaly
which can be written as
\bea
\label{eq:ax_wilson}
 \anx &=&  \frac{r_w}{2} \sum_{\mu} \Bigl\{
    4 \ \tr [ D_w^{-1} (x,x) \gm5 ]
    - \ \tr [ D_w^{-1}(x,x+\hat\mu) \gm5 U_{\mu}(x) ]  \nn
&& - \ \tr [ D_w^{-1}(x+\hat\mu,x) \gm5 U_{\mu}^{\dagger}(x) ]
   - \ \tr [ D_w^{-1}(x-\hat\mu,x) \gm5 U_{\mu}(x-\hat\mu) ] \nn
&& - \ \tr [ D_w^{-1}(x,x-\hat\mu) \gm5 U_{\mu}^{\dagger}(x-\hat\mu)] \ \Bigr\}.
\eea
In the classical continuum limit ( $ a \to 0 $ ), (\ref{eq:ax_wilson})
was shown \cite{karsten81}-\cite{fuji84} to agree with the
Chern-Pontryagin density. On the other hand, the sum of (\ref{eq:ax_wilson})
over all sites on a finite lattice must be zero in the massless limit,
as shown in Eq. (\ref{eq:ax_zero}), since $ D_w $ does not have exact zero
modes. So, in general, the axial anomaly (\ref{eq:ax_wilson}) of the
Wilson-Dirac fermion operator on a finite lattice must be different from
the Chern-Pontryagin density. Only in the classical continuum limit when
the lattice spacing is equal to zero and the number of sites is infinite,
then the exact zero modes could possibly emerge and the correct axial
anomaly might be recovered. If this does happen, then we would expect that
in the massless limit, the sum of (\ref{eq:ax_wilson}) over all sites must
undergo a discontinuous transition around the point $ a = 0 $,
for topologically nontrivial gauge background.

Next, we consider a {\em topologically proper} Dirac operator
which has exact zero modes satisfying the Atiyah-Singer index theorem.
Suppose that one has constructed a lattice Dirac operator\footnote{ This
is a generalized Neuberger-Dirac operator. } as the following
\beq
\label{eq:D01}
D = 5 \left( \sqrt{ D_w^{\dagger} D_w } + D_w \right)
       \left( 3 \sqrt{ D_w^{\dagger} D_w } - 2 D_w \right)^{-1}
\eeq
where $ D_w $ is the Wilson-Dirac operator with a negative
mass parameter $ m_0 $ and Wilson parameter $ r_w $,
\bea
\label{eq:Dw_m0}
D_w = - m_0
      + \frac{1}{2} [ \gamma_{\mu} ( \nabstar{\mu} + \nab{\mu} ) -
                      r_w \nabstar{\mu} \nab{\mu} ]
\eea
For the moment, we fix $ m_0 = 1 $ and $ r_w = 1 $.
Since $ D $ satisfies the Ginsparg-Wilson relation (\ref{eq:gwr}),
the anomaly function of $ D $ can be easily obtained from
Eq. (\ref{eq:ax_anomaly}).
First, we investigate the properties of $ D $ [ Eq. (\ref{eq:D01}) ] in
background gauge fields with constant field tensor. As we expect,
$ D $ possesses exact zero modes satisfying the Atiyah-Singer index theorem.
However, the anomaly function is not constant, though the fluctuations are
not very large. The deviation of the anomaly function of $ D $ from the
Chern-Pontryagin density is defined as
\bea
\label{eq:dev}
\delta_D \equiv \frac{1}{N_s} \sum_{x} \frac{| \anx - \ancx | }{ | \ancx | }
\eea
where the summation runs over all sites, and $ N_s $ is the total number
of sites, and $ \ancx $ is the Chern-Pontryagin density which
is equal to
\bea
\label{eq:ancx}
\ancx = \frac{1}{ 2 \pi } \epsilon_{\mu\nu} F_{\mu\nu} (x)
\eea
on a torus. For constant field tensor with topological charge $ Q $,
it becomes a constant
\bea
\label{eq:ancc}
\ancx = \frac{ 2 Q }{ L_1 L_2 }
\eea
In Table 1, we list the deviation of the anomaly function of $ D $,
$ \delta_D $, for topological charge $ Q = 1 $ to $ Q = 8 $
on a $ 12 \times 12 $ lattice.
Although the deviation $ \delta_D $ in the third columm seems to be
tiny in each case, it should be regarded as large when comparing
with the deviation $ \delta_{D'} $ of a local $ D' $
[ Eq. (\ref{eq:D06}) ] listed in the fourth coluum. In fact, the
fluctuations of the axial anomaly of $ D $ can be seen clearly by plotting
$ \anx $ versus each site of the lattice, as shown in Fig. 1.
The horizontal line at height $ 1/72 $ denotes the constant
Chern-Pontryagin density, $ \ancx $ for $ Q = 1 $ on a $ 12 \times 12 $
torus. The triangles denote the axial anomaly of $ D $,
and its fluctuations are due to the non-locality of $ D $.
The average of the axial anomaly of $ D $, of course, must agree
with the Chern-Pontryagin density, since its index is equal to $ Q $.
On the other hand, the axial anomaly [ Eq. (\ref{eq:ax_wilson}) ] of the
standard Wilson-Dirac fermion operator $ D_w $ [ Eq. (\ref{eq:Dw}) ] with
$ m_f = 0 $ and $ r_w = 1 $, which is plotted as squares in Fig. 1,
is very different from the Chern-Pontryagin density at each site.
This is due to the fact that $ D_w $ does not have exact zero modes, thus
the sum of its axial anomaly over all sites is zero on any finite
lattice. So, the deviation of the axial anomaly of massless Wilson-Dirac
operator, $ \delta_{D_w} $, is equal to one for any $ Q $, as listed in the
last coluum of Table 1.

{\footnotesize
\begin{table}
\begin{center}
\begin{tabular}{|c|c|c|c|c|}
\hline
 Q  &  $ \ancx = 2 Q/(L_1 L_2) $
    &  $ \delta_D $ & $ \delta_{D'} $ & $ \delta_{D_w}  $ \\
\hline
\hline
 1  &  $ 1.389 \times 10^{-2} $ &  $ 6.232 \times 10^{-2} $ &
       $ 2.687 \times 10^{-4} $ & 1.0 \\
\hline
 2  &  $ 2.778 \times 10^{-2} $ &  $ 2.434 \times 10^{-2} $ &
       $ 2.413 \times 10^{-4} $ & 1.0 \\
\hline
 3  &  $ 4.167 \times 10^{-2} $ &  $ 8.026 \times 10^{-3} $ &
       $ 2.036 \times 10^{-4} $ & 1.0 \\
\hline
 4  &  $ 5.556 \times 10^{-2} $ &  $ 3.192 \times 10^{-3} $ &
       $ 1.982 \times 10^{-4} $ & 1.0 \\
\hline
 5  &  $ 6.944 \times 10^{-2} $ &  $ 1.026 \times 10^{-3} $ &
       $ 1.363 \times 10^{-4} $ & 1.0 \\
\hline
 6  &  $ 8.333 \times 10^{-2} $ &  $ 5.149 \times 10^{-4} $ &
       $ 1.184 \times 10^{-4} $ & 1.0 \\
\hline
 7  &  $ 9.722 \times 10^{-2} $ &  $ 1.910 \times 10^{-4} $ &
       $ 5.873 \times 10^{-5} $ & 1.0 \\
\hline
 8  &  $ 1.111 \times 10^{-1} $ &  $ 1.061 \times 10^{-4} $ &
       $ 3.446 \times 10^{-5} $ & 1.0 \\
\hline
\end{tabular}
\end{center}
\caption{The chiral anomaly versus the topological charge $ Q $.
The exact solution is listed in the second coluum.
The deviations of the chiral anomaly
for $ D $ [ Eq. (\ref{eq:D01}) ] and
$ D' $ [ Eq. (\ref{eq:D06}) ]
are listed in the third and fourth coluums
respectively. The last column lists the deviations for the
Wilson-Dirac fermion.}
\label{table:anx}
\end{table}
}

\psfigure 5.0in -0.2in {fig:anomaly_cons} {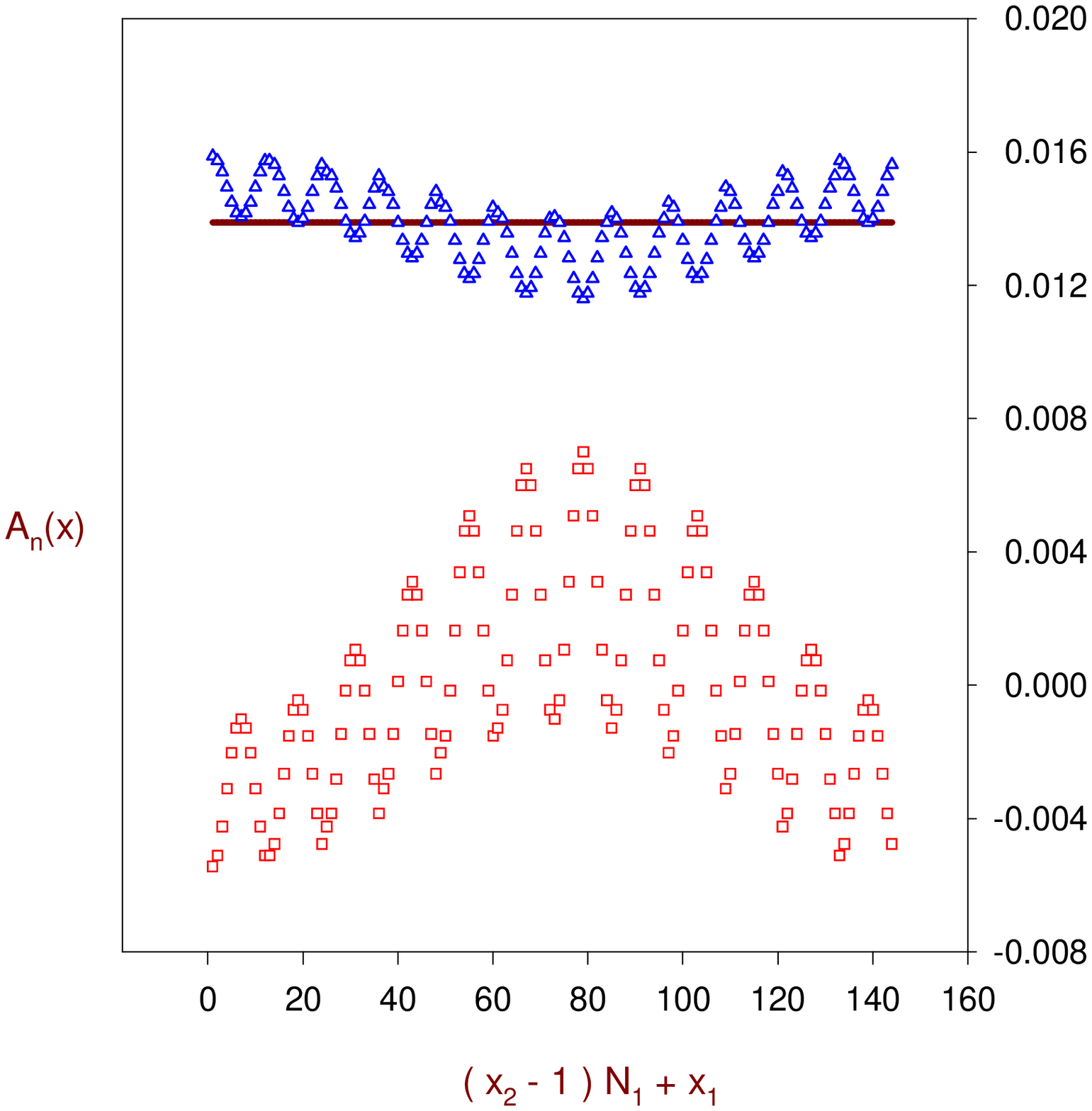} {
The anomaly function on a $ 12 \times 12 $ lattice.
The background gauge field has constant field tensor with
topological charge $ Q = 1 $. The horizontal line at height $ 1/72 $
is the Chern-Pontryagin density on the torus. The axial anomaly
of $ D $ [ Eq. (\ref{eq:D01}) ] is denoted by triangles which fluctuate
around the exact solution, while that of the Wilson-Dirac operator
[ Eq. (\ref{eq:Dw}) with $ m_f = 0 $ and $ r_w = 1 $ ] by
squares which are far away from the exact solution.
The axial anomaly of $ D' $
[ Eq. (\ref{eq:D06}) ] coincides with the exact solution
( the horizontal line at height $ 1/72 $ ). }

The non-locality of $ D $ can be easily seen by plotting
$ D(x,y) $ as a function of $ | x - y | $.
In Fig. 2, one of the Dirac components of $ | D(x,0) | $, say,
$ | D_{11}(x,0) | $ is plotted versus the distance $ | x | $,
on a $ 12 \times 12 $ lattice with periodic boundary conditions,
in a constant background gauge field with topological charge $ Q = 1 $.
It is obvious that $ D(x,y) $ is nonlocal.

\psfigure 5.0in -0.2in {fig:locality} {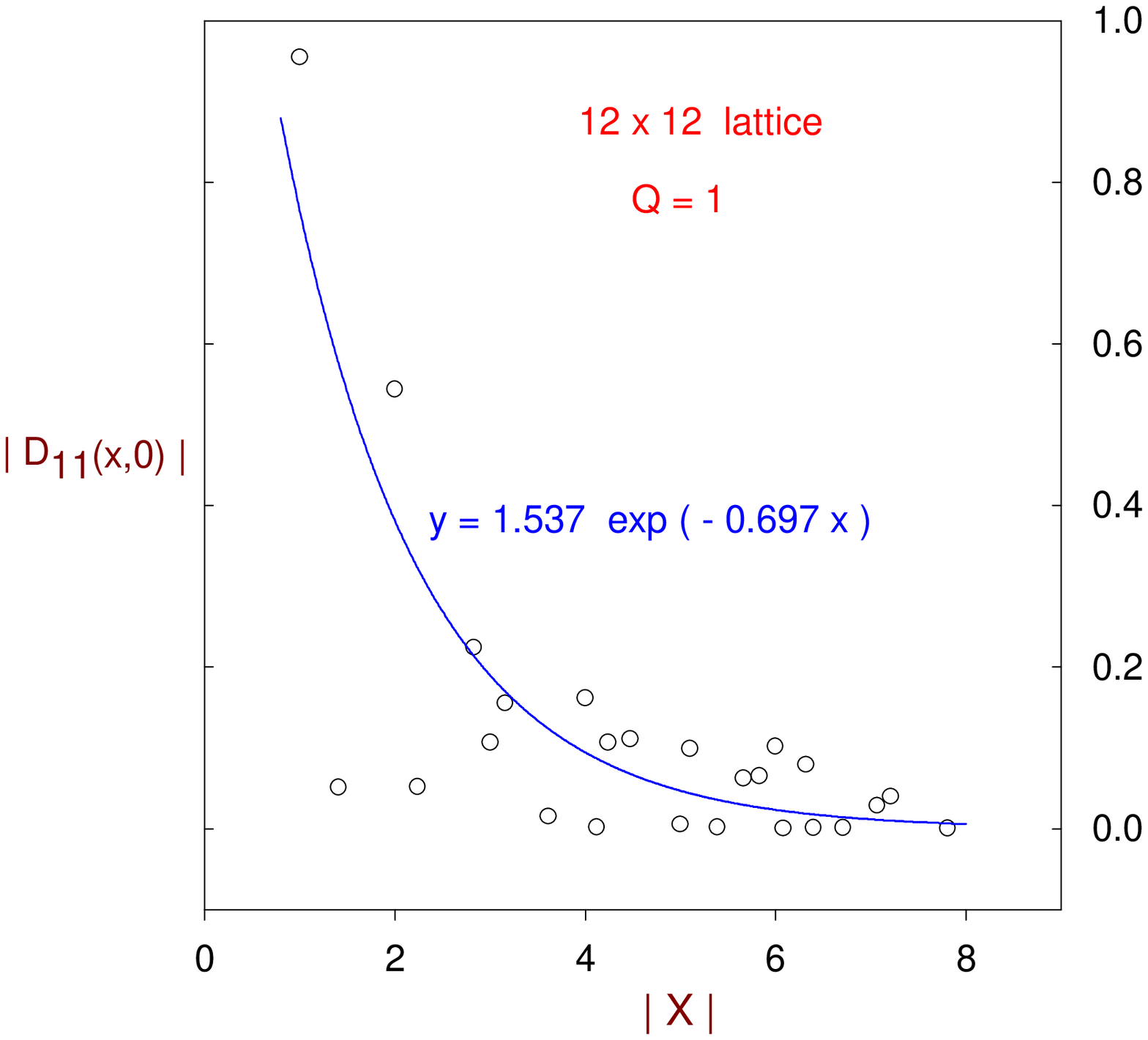} {
One of the Dirac components of $ D(x,0) $, $ | D_{11} (x,0) | $, is plotted
as a function of $ | x | $. The lattice is $ 12 \times 12 $ with periodic
boundary conditions. The background gauge field has constant field
tensor with topological charge $ Q = 1 $. All data points with the
same distance $ | x | $ from the origin have been averaged.
The solid line is an exponential fit to the data points. }

Although $ D $ is topologically proper, the non-locality of $ D $
must yield incorrect chiral anomaly and fermion determinant.
In Table 2, we compute the fermion determinant of $ D $
for different topological sectors of constant field tensor,
and compare them with the exact solution on a torus \cite{sachs_wipf},
\bea
\label{eq:exactD}
  \det_1[ D (Q) ]_{exact}
= N \sqrt{ \left( \frac{ L_1 L_2 }{ 2 |Q| } \right)^{|Q|} }
\eea
where the zero modes have been omitted and
the normalization constant is fixed by $ N = \sqrt{ \frac{2}{L_1 L2} } $
such that $ \det_1[ D(1) ]_{exact} = 1 $.  It is evident that the
fermion determinant of $ D $ ( in the third coluum ) disagrees with
the exact solution ( in the second coluum ), while those of a local
$ D' $ [ Eq. (\ref{eq:D06}) ] ( in the fourth coluum ) agree fairly well
with the exact solutions. The fermion determinants for the Wilson-Dirac
operator are listed in the fifth coluum, where all eigenvalues of $ D_w $
are taken into account. Thus its fermion determinant must become
smaller for larger $ Q $ due to the emergence of small real eigenvalues.
If we discard these small real eigenvalues of $ D_w $, which correspond
to the exact zero modes of $ D $, then the resulting fermion determinant of
$ D_w $ is listed in the last coluum of Table 2, but still in disagreement
with the exact solution. In general, the fermion determinant ratio
of $ D_w $ between any two gauge configurations ( no matter within the
same topological sector, or in two different sectors ) disagrees with the
exact solution, except for the trivial sector with $ Q = 0 $.

{\footnotesize
\begin{table}
\begin{center}
\begin{tabular}{|c|c|c|c|c|c|}
\hline
 Q  &   $ \det_{1}[D(Q)]_{exact} $  &  $ \det_{1}[D(Q)] $
    &   $ \det_{1}[D'(Q)] $  &   $  \det[D_w(Q)] $ &  $  \det'[D_w(Q)] $ \\
\hline
\hline
  1   &   1.00000   &   1.00000  &    1.00000  &    1.00000  &   1.00000 \\
\hline
  2   &   4.24264   &   5.21339  &    4.22458  &    0.34182  &   3.98005 \\
\hline
  3   &   13.8564   &   23.5281  &    13.7284  &    0.14620  &   11.8619 \\
\hline
  4   &   38.1838   &   98.6168  &    37.2195  &    0.07092  &   28.8108 \\
\hline
  5   &   92.7342   &   394.114  &    90.3215  &    0.03841  &   61.1240 \\
\hline
  6   &   203.647   &   1535.83  &    198.435  &    0.02185  &   112.466 \\
\hline
  7   &   411.296   &   5759.04  &    391.581  &    0.01337  &   190.058 \\
\hline
  8   &   773.221   &   21275.2  &    726.985  &    0.00854  &   293.285 \\
\hline
\end{tabular}
\end{center}
\caption{The fermion determinant versus the topological charge $ Q $.
The exact solution for a torus of size $ 12 \times 12 $ is listed in
the second coluum, which is computed according to Eq. (\ref{eq:exactD}).
The normalization constants are fixed by the first row in each case.
The lattice Dirac operators $ D $, $ D' $ and $ D_w $ are defined in
Eqs. (\ref{eq:D01}), (\ref{eq:D06}) and (\ref{eq:Dw}) respectively. }
\label{table:det}
\end{table}
}

As long as we have a topologically proper $ D $, we can use the
topologically invariant transformation (\ref{eq:twc}) to obtain a local
$ D' $ such that the anomaly function and the fermion determinant both
agree with the exact solutions for smooth gauge configurations.
For simplicity, we fix $ R = r \Id $ in (\ref{eq:twc}).
Note that there is a large range of $ r $ values to render $ D' $ local
for constant background gauge fields. For instance, we can pick $ r = 0.5 $,
then (\ref{eq:twc}) gives
\bea
\label{eq:D06}
D' = 10 \left( \sqrt{ D_w^{\dagger} D_w } + D_w \right)
       \left( 11 \sqrt{ D_w^{\dagger} D_w } + D_w \right)^{-1}
\eea
It is easy to check that $ D' $ is local by plotting $ | D'(x,y) | $
versus $ | x - y | $, as shown in Fig. 3. The chiral anomaly and fermion
determinant of $ D' $ agree very well with the exact solutions, as listed
in Table 1 and Table 2.

Now we can introduce local fluctuations to the field tensor through
the sinusoidal terms in Eqs. (\ref{eq:A1_2d}) and (\ref{eq:A2_2d}).
For $ Q = 1 $, $ h_1 = 0.1 $, $ h_2 = 0.2 $, $ A_1^{(0)} = 0.3 $,
$ A_2^{(0)} = 0.4 $ and $ n_1 = n_2 = 1 $ on a $ 12 \times 12 $ lattice,
the anomaly function of $ D' $ is computed and compared with
the Chern-Pontryagin density on the torus, as shown in Fig. \ref{fig:an12}.
It is evident that $ A_n(x) $ agress with $ A_c(x) $ very well
at each site.

\psfigure 5.0in -0.2in {fig:local2} {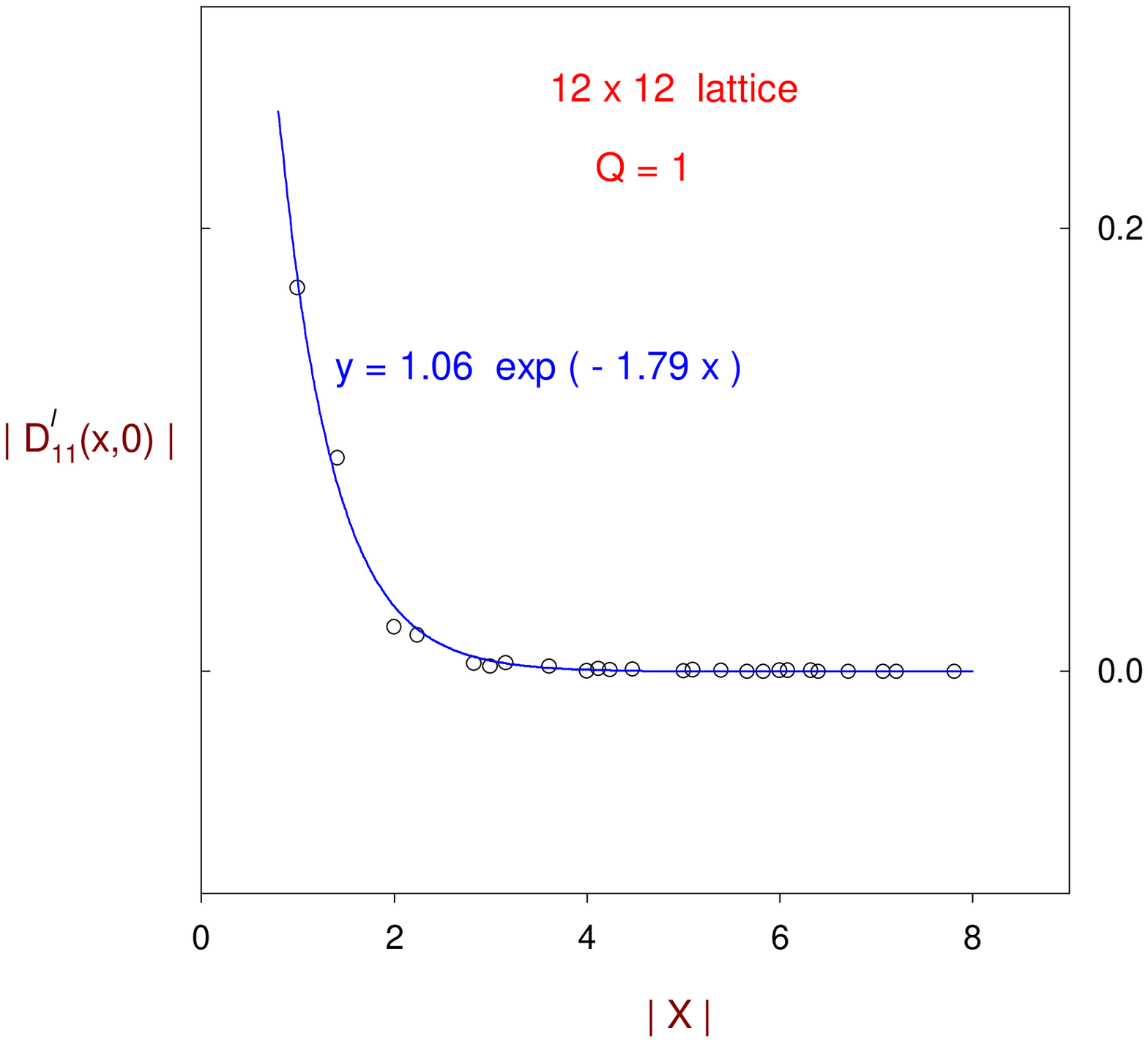} {
$ | D^{'}_{11} (x,0) | $ is plotted
as a function of $ | x | $. The lattice is $ 12 \times 12 $ with periodic
boundary conditions. The constant background gauge field has
topological charge $ Q = 1 $. All data points at the
same distance $ | x | $ from the origin have been averaged.
The solid line is an exponential fit to the data points.
The same decay constant also fits very well for all other Dirac components
of $ D(x,y) $ and for any reference point $ y $. }

\psfigure 5.0in -0.2in {fig:an12} {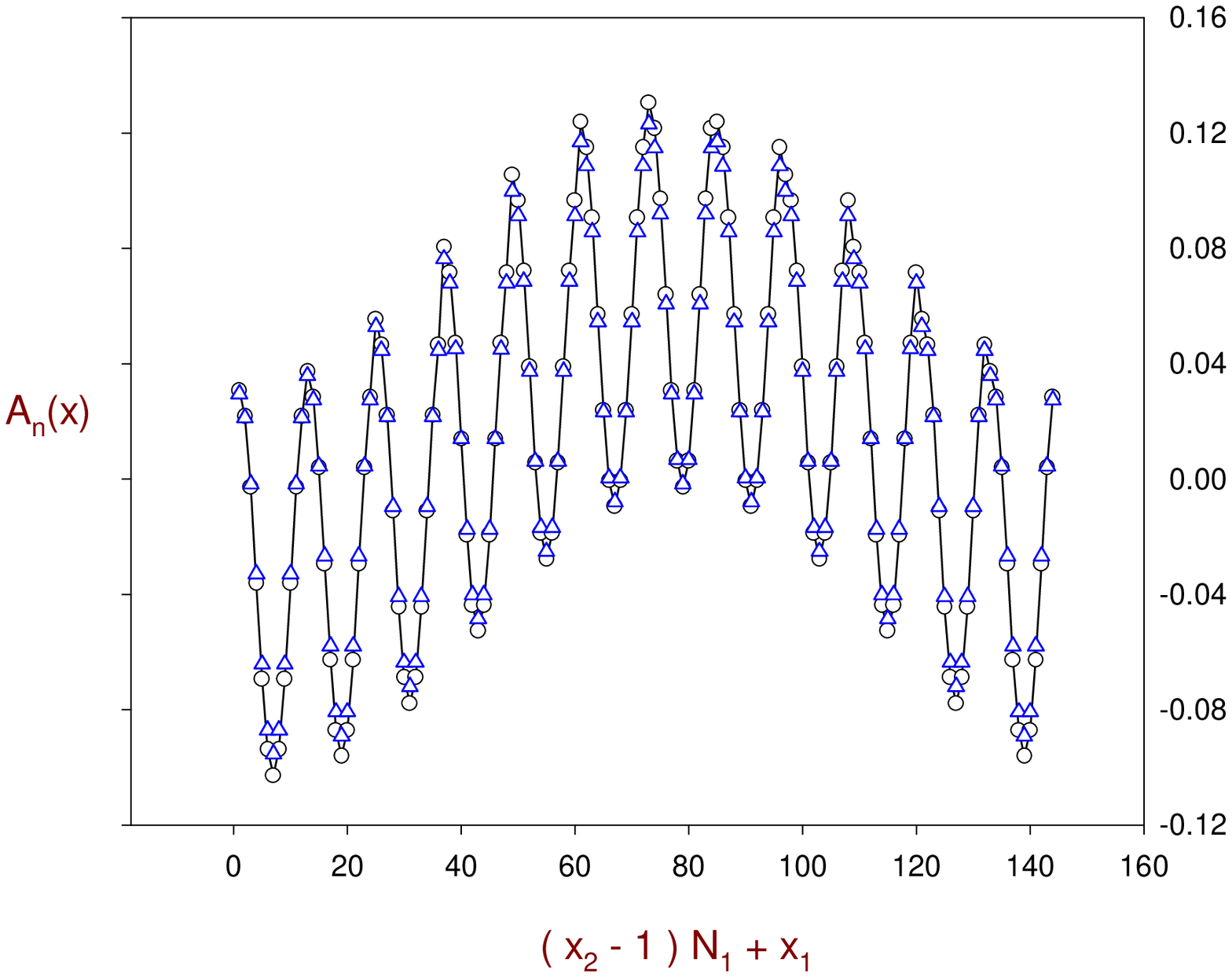} {
The anomaly function of $ D' $ [ Eq. (\ref{eq:D06}) ]
on a $ 12 \times 12 $ lattice versus the Chern-Pontryagin density
$ \ancx $ [ Eq. (\ref{eq:ancx}) ] on the torus.
The parameters of the background gauge field are $ Q = 1 $,
$ h_1 = 0.1 $, $ h_2 = 0.2 $,
$ A_1^{(0)} = 0.3 $, $ A_2^{(0)} = 0.4 $ and $ n_1 = n_2 = 1 $.
The Chern-Pontryagin density is denoted by circles which are
connected by line segments for the visual purpose.
The anomaly function of $ D' $ is denoted by triangles.
}

Next we consider another lattice Dirac operator $ D'' $ which has the
same form of $ D' $ (\ref{eq:D06}) except $ m_0 = 0.1 $ in the $ D_w $,
\bea
\label{eq:D''}
D'' = 10 \left( \sqrt{ D_w^{\dagger} D_w } + D_w \right)
       \left( 11 \sqrt{ D_w^{\dagger} D_w } + D_w \right)^{-1} \\
\eea
where
$$
D_w = - 0.1
      + \frac{1}{2} [ \gamma_{\mu} ( \nabstar{\mu} + \nab{\mu} ) -
                      \nabstar{\mu} \nab{\mu} ] \ .
$$
Again we investigate the properties of $ D'' $ in a background gauge field
of constant field tensor. Then we find
that $ D'' $ does not have any zero modes for $ Q \ge 5 $, in contrast
to the behaviors of $ D $ and $ D' $ having zero modes satisfying the
Atiyah-Singer index theorem. In terms of the general index theorem,
Eq. (\ref{eq:cD}), we list the topological characteristics
of $ D $, $ D' $ and $ D'' $ in Table \ref{table:cD}.
It shows clearly that $ c[D''] $ depends on the gauge configuration
through $ D'' $. Note that $ D $ is {\em nonlocal} for $ Q = 1 $ to $ Q = 8 $,
but {\em the Atiyah-Singer index theorem is still satisfied}. We recall that
{\em the nonlocal $ D $ and the local $ D' $ are related by the topological
invariant transformation (\ref{eq:twc}) which preserves the zero modes
as well as the index}.

{\footnotesize
\begin{table}
\begin{center}
\begin{tabular}{|c|c|c|c|}
\hline
 Q  &  $ c[D] $  &  $ c[D'] $ &  $ c[D''] $  \\
\hline
\hline
 1  &  1 &  1 & 1   \\
\hline
 2  &  1 &  1 & 1   \\
\hline
 3  &  1 &  1 & 1   \\
\hline
 4  &  1 &  1 & 1   \\
\hline
 5  &  1 &  1 & 0   \\
\hline
 6  &  1 &  1 & 0   \\
\hline
 7  &  1 &  1 & 0   \\
\hline
 8  &  1 &  1 & 0   \\
\hline
\end{tabular}
\end{center}
\caption{The topological characteristics of
$ D $, $ D' $ and $ D'' $ versus the topological charge $ Q $
on a $ 12 \times 12 $ lattice.}
\label{table:cD}
\end{table}
}

Now we introduce local fluctuations to the field tensor through
the sinusoidal term in Eq. (\ref{eq:A1_2d}). Starting with constant field
tensor at $ Q = 4 $, we gradually increase the amplitude of the
sinusoidal term, $ A_1^{(0)} $ with $ n_2 = 1 $.
The results are shown in Table \ref{table:cD2}.
For $ A_1^{(0)} \le 0.0802 $, the index of $ D'' $ is 4, satisfying
the Atiyah-Singer index theorem. However, around $ A_1^{(0)} = 0.0803 $,
the index of $ D'' $ becomes $ 3 $, thus
the index theorem (\ref{eq:cD}) is violated for any integer $ c[D''] $.
Therefore, we deduce that, in general, if $ c[D] $ is not a constant
for any integer topological charge $ Q $, we can generalize $ c[D] $ to a
rational number, then the index theorem (\ref{eq:cD}) also holds for
{\em any} rough gauge fields.
It is remarkable that the index of $ D'' $ changes so abruptly around
$ A_1^{(0)} = 0.0802 \sim 0.0803 $, a singular behavior characterizes
the topological phase transition. It is very unimaginable for us to
see odd number of zero modes ( $ N_{-} = 3 $ ) in a
background gauge field of even number of topological charges ( $ Q = 4 $ ).
The underlying nonperturbative mechanism is very intriguing, and
might have many unexpected consequences.

We note in passing that if we substitute $ D' $ [ Eq. (\ref{eq:D06}) ]
and $ D'' $ [ Eq. (\ref{eq:D''}) ] by the Neuberger-Dirac operator
$ D_h = \Id + D_w ( D_w^{\dagger} D_w )^{-1/2} $ with $ m_0 = 1 $
and $ m_0 = 0.1 $ respectively. The essential features of above numerical
results are almost the same. In particular, the good agreement
between the anomaly function and the Chern-Pontryagin density
at each site, as shown in Fig. 4. We also note that the geometrical
aspects of chiral anomalies in the overlap was investigated
by Neuberger \cite{hn98:3}.

{\footnotesize
\begin{table}
\begin{center}
\begin{tabular}{|c|c|c|c|}
\hline
 Q  &  $ A_1^{(0)} $  &  $ N_{-} - N_{+} $ &  $ c[D''] $  \\
\hline
\hline
 4  &  0.0100 &  4 & 1   \\
\hline
 4  &  0.0200 &  4 & 1   \\
\hline
 4  &  0.0400 &  4 & 1   \\
\hline
 4  &  0.0600 &  4 & 1   \\
\hline
 4  &  0.0800 &  4 & 1   \\
\hline
 4  &  0.0801 &  4 & 1   \\
\hline
 4  &  0.0802 &  4 & 1   \\
\hline
 4  &  0.0803 &  3 & 3/4  \\
\hline
 4  &  0.0804 &  3 & 3/4  \\
\hline
 4  &  0.0810 &  3 & 3/4  \\
\hline
 4  &  0.0820 &  3 & 3/4  \\
\hline
 4  &  0.0900 &  3 & 3/4  \\
\hline
\end{tabular}
\end{center}
\caption{The topological characteristics of
$ D'' $ versus the amplitude of the local sinusoidal fluctuation
on a $ 12 \times 12 $ lattice, where $ c[D''] $ is obtained in accordance
with the general index theorem (\ref{eq:cD}). }
\label{table:cD2}
\end{table}
}

\section{ Summary and Discussions }

In summary, we have shown that the condition for lattice Dirac
operator $ D $ ( or $ D'= D (\Id + R D )^{-1} $ ) to reproduce the
continuum anomaly function at each site of a finite lattice for
smooth background gauge fields is that $ D $ possesses exact zero modes
satisfying the Atiyah-Singer index theorem
( i.e., $ D $ is {\em topologically proper} ).
This is the basic requirement for any Dirac operator to be used
for lattice QCD if one wishes to reproduce correct chiral anomaly
and fermion determinant ( ratio ) on a finite lattice.
It is evident that the standard Wilson-Dirac fermion operator does
{\em not} meet this basic requirement.
If $ D $ is topologically proper but nonlocal, then one can try to use
(\ref{eq:twc}) to transform $ D $ into a local $ D'= D ( \Id + R D )^{-1} $
such that the anomaly function of $ D' $ agrees with the Chern-Pontryagin
density in continuum. Note that there is no need for fine-tunings of $ R $.
The set of transformations $ \{ \CT(R) \} $ defined in (\ref{eq:twc})
form an abelian group with the group parameter space $\{ R \}$.
It is evident that (\ref{eq:twc}) is one of the simplest topologically
invariant transformations which preserve the zero modes and the index
of $ D $.

The chiral anomaly on the lattice can be understood from two seemingly
different viewpoints. In general, if we have lattice Dirac operator
$ D $ which breaks the usual chiral symmetry according to Eq. (\ref{eq:DB}),
then the chiral anomaly can be obtained through the chiral symmetry breaking
term [ Eq. (\ref{eq:ax_anomaly} ], while the fermion measure is invariant
under the usual chiral transformation, in contrast to the continuum theory
in which the non-invariance of the fermion measure under the chiral
transformation leads to the emergence of chiral anomaly
after regularization \cite{fuji79}.
Another viewpoint to understand the chiral anomaly on a finite lattice is
only for those $ D $ having chirally symmetric limit $ D_c $ under the
transformation (\ref{eq:twc}). Then $ D $ satisfies the Ginsparg-Wilson
relation (\ref{eq:gwr}) which is the lattice chiral symmtry,
thus the lattice action is invariant under the lattice chiral transformation.
However, the fermionic measure is not invariant
under the lattice chiral transformation, and thus leads to the
chiral anomaly \cite{ml98:2}, reminiscent of the situation in continuum.
Therefore, for Ginsparg-Wilson Dirac operator, one must obtain
the same anomaly function (\ref{eq:anx_gw}) no matter which viewpoint
one prefers. However, Eqs. (\ref{eq:ax_anomaly}) and (\ref{eq:anx}) are
the general formulas of the anomaly function for {\em any} lattice Dirac
operator.

There is a vital difference between the continuum Dirac operator
and the lattice ones.
In continuum, the Dirac operator in a smooth non-trivial
gauge background must have exact zero modes satisfying the Atiyah-Singer
index theorem. On the other hand, lattice Dirac operator does not
necessarily possess exact zero modes in a smooth non-trivial gauge
background, not to mention satisfying the Atiyah-Singer index theorem.
This naturally leads to the concept of topological characteristics,
$ c[D] $, which is an intrinsic attribute of $ D $, in general, is
{\em not} entirely due to the non-locality of $ D $ and/or the occurrence
of species doubling. Even if one starts with a lattice Dirac operator
$ D $ which is local and free of species doubling in a constant background
gauge field, and the zero modes of $ D $ also satisfies the Atiyah-Singer
index theorem. However, as we increase
the topological charge or the local fluctuations of the background gauge
field, $ D $ may undergo a topological phase transition
and its zero modes may violate the Atiyah-Singer index theorem.
In fact $ D $ has become nonlocal long before it reaches the topological
phase transition point. So, the non-locality of $ D $ cannot be the crucial
factor for violations of the Atiyah-Singer index theorem, though it explains
the breakdown of the anomaly function (\ref{eq:anxc}) unambiguously.
If one tries to explain the violations of Atiyah-Singer index theorem
in terms of species doubling, then one has great difficulties
to explain the occurrence of odd number of zero modes in a gauge background
of even number of topological charges, as shown in Table \ref{table:cD2}.
Therefore we conclude that the topological characteristics of lattice
Dirac operator $ D $ is a basic attribute of
$ D $, which in general is a rational number appearing in the general index
theorem (\ref{eq:cD}) for integer topological charge $ Q $.

Given a lattice Dirac operator $ D $, we cannot gaurantee that for any
gauge configuration, there exists an operator $ R $ such that the
topologically invariant transformation (\ref{eq:twc}) gives a {\em local}
and {\em smooth} $ D' $. The notion of locality is usually defined
as follows :
$ | D(x,y) | \sim \exp ( - M |x-y|/a ) $ with $ M = O(1) $, or
$ | D(x,y) | = 0 $ for $ | x - y | > d $ with $ d = O(a) $.
Although $ | D(x,y) | $ can be fitted by
$ \exp ( - M |x-y|/a ) $ with $ M = O(1) $, it may turn out that
$ | D(x,y) | $ is still highly fluctuating and/or anisotropic.
So, it is necessary to introduce the notion of "smoothness" which
measures the fluctuations of $ | D(x,y) | $ with respect to
$ \exp ( - M |x-y|/a ) $.
To determine the criterion for the gauge configuration such that $ D'$
is local and smooth is beyond the scope of the present paper. It is
apparent that such criterion should take into account of the
fluctuations between the plaquettes as well as the deviations of each
plaquette from the identity. Otherwise the criterion is only useful
in the weak field limit. For example, the Neuberger-Dirac operator $ D_h $
with $ m_0 = r_w = 1 $ and $ D' $ [ Eq. (\ref{eq:D06}) ] are both local
and smooth for gauge configurations with constant field tensors up
to very large $ Q $ values ( see Table 3 ),
though each plaquette is very different from the identity.
If one takes into account of the degree of freedom provided by the
transformation (\ref{eq:twc}), then the criterion for a smooth
gauge configuration could be less stringent than without it.

A scheme for dynamical fermion simulation of lattice QCD is
outlined in Section 3. Since the direct computation of fermion determinant
is prohibitively expensive for lattice QCD, a practical implementation
of the scheme must be worked out before we can proceed to
any realistic calculations. Nevertheless, the prescription
of decomposing link configurations into distinct topological
sectors and performing Monte Carlo simulations for each relevant
topological sector individually may open new avenues to tackle the
longstanding problems in QCD.

\appendix

\section{ }

In this appendix, we give an example of prescribed background gauge field
which has been used in ref. \cite{twc98:4,twc98:10a} for two dimensional
and four dimensional lattices respectively.
On a 4-dimensional torus ( $ x_{\mu} \in [0,L_{\mu}], \mu = 1, \cdots, 4 $ ),
the simplest nontrivial gauge fields can be represented as
\bea
\label{eq:A1}
A_1(x) &=&  t^{a} \left[  \frac{ 2 \pi h_1 }{L_1}
                        - \frac{ 2 \pi q_1 x_2 }{ L_1 L_2 }
     +  A_1^{(0)} \sin \left( \frac{ 2 \pi n_2 }{L_2} x_2 \right) \right]  \\
\label{eq:A2}
A_2(x) &=&  t^{a} \left[ \frac{ 2 \pi h_2 }{L_2}
     +  A_2^{(0)} \sin \left( \frac{ 2 \pi n_1 }{L_1} x_1 \right) \right] \\
\label{eq:A3}
A_3(x) &=&  t^{a} \left[ \frac{ 2 \pi h_3 }{L_3}
               - \frac{ 2 \pi q_2 x_4 }{ L_3 L_4 }
     +  A_3^{(0)} \sin \left( \frac{ 2 \pi n_4 }{L_4} x_4 \right) \right] \\
\label{eq:A4}
A_4(x) &=&  t^{a} \left[ \frac{ 2 \pi h_4 }{L_4}
     +  A_4^{(0)} \sin \left( \frac{ 2 \pi n_3 }{L_3} x_3 \right) \right]
\eea
where $ t^a $ is any one of the generators of the gauge group with the
normalization $ \tr( t^a t^b ) = n \ \delta^{ab} $,
and $ q_1 $ and $ q_2 $ are integers.
The global part is characterized by the topological charge
\beq
Q = \frac{1}{32\pi^2} \int d^4 x \ \epsilon_{\mu\nu\lambda\sigma} \
    \tr( F_{\mu\nu} F_{\lambda\sigma} ) = n \ q_1 \ q_2
\label{eq:ntop}
\eeq
which is an integer. The harmonic parts are parameterized by four
real constants $ h_1 $, $ h_2 $, $ h_3 $ and $ h_4 $.
The local parts are chosen to be sinusoidal
fluctuations with amplitudes $ A_1^{(0)} $, $ A_2^{(0)} $, $ A_3^{(0)} $
and $ A_4^{(0)} $, and
frequencies $ \frac{ 2 \pi n_2 }{L_2} $, $ \frac{ 2 \pi n_1 }{L_1} $,
$ \frac{ 2 \pi n_4 }{L_4} $ and $ \frac{ 2 \pi n_3 }{L_3} $ where
$ n_1 $, $ n_2 $, $ n_3 $ and $ n_4 $  are integers.
The discontinuity of $ A_1(x) $ ( $ A_3(x) $ ) at $ x_2 = L_2 $
( $ x_4 = L_4 $ ) due to the global part
only amounts to a gauge transformation.
The field tensors $ F_{12} $ and $ F_{34} $ are continuous
on the torus, while other $ F's $ are zero.

To transcribe the continuum gauge fields to the lattice, we take the lattice
sites at $ x_\mu = 0, a, ..., ( N_\mu - 1 ) a $, where $ a $ is the lattice
spacing and $ L_\mu = N_\mu a $ is the lattice size.
Then the link variables are
\bea
\label{eq:U1}
U_1(x) &=& \exp \left[ \text{i} A_1(x) a \right] \\
\label{eq:U2}
U_2(x) &=& \exp \left[ \text{i} A_2(x) a
 + \text{i} \delta_{x_2,(N_2 - 1)a} \frac{ 2 \pi q_1 x_1 }{L_1} \ t^a \right] \\
\label{eq:U3}
U_3(x) &=& \exp \left[ \text{i} A_3(x) a \right] \\
\label{eq:U4}
U_4(x) &=& \exp \left[ \text{i} A_4(x) a
 + \text{i} \delta_{x_4,(N_4 - 1)a} \frac{ 2 \pi q_2 x_3 }{L_3} \ t^a \right]
\eea
The last term in the exponent of $ U_2(x) $ ( $ U_4(x) $ ) is included to
ensure that the field tensor $ F_{12} $ ( $ F_{34} $ ) which is defined by
the ordered product of link variables around a plaquette is continuous on
the torus.
For $ U(1) $ gauge field on a two-dimensional lattice, the link
variabes are reduced to Eqs. (\ref{eq:U1}) and (\ref{eq:U2}) with
\bea
\label{eq:A1_2d}
A_1(x) &=&  \frac{ 2 \pi h_1 }{L_1}
                        - \frac{ 2 \pi Q x_2 }{ L_1 L_2 }
     +  A_1^{(0)} \sin \left( \frac{ 2 \pi n_2 }{L_2} x_2 \right) \\
\label{eq:A2_2d}
A_2(x) &=&  \frac{ 2 \pi h_2 }{L_2}
     +  A_2^{(0)} \sin \left( \frac{ 2 \pi n_1 }{L_1} x_1 \right)
\eea
where the topological charge is
\bea
\label{eq:Q_2d}
Q = \frac{1}{2 \pi} \int d^2 x \ F_{12}
\eea

\bigskip
\bigskip

\flushpar
{\bf Acknowledgement }
\bigskip

\noindent
This work was supported by the National Science Council, R.O.C.
under the grant number NSC88-2112-M002-016.

\vfill\eject

\vfill\eject


\begin{thebibliography}{15}

\bibitem{ABJ} S. Adler, Phys. Rev. 177 (1969) 2426; J.S. Bell and R. Jackiw,
Nuovo Cimento 60 (1969) 47.

\bibitem{tHooft76} G 't Hooft, Phys. Rev. Lett. 37 (1976) 8;
Phys. Rev. D17 (1976) 3432; Phys. Rep. 142 (1986) 357.

\bibitem{weinberg96} For a detailed account and references, see, e.g.,
S. Weinberg, {\em The Quantum Theory of Fields}, vol. II,
( Cambridge University Press, 1996).

\bibitem{twc98:6a} T.W. Chiu and S.V. Zenkin, Phys. Rev. D59, 074501 (1999).

\bibitem{wilson75} K. G. Wilson, in {\em New Phenomena in Subnuclear Physics},
proceedings of the 14th course of the International School of Subnuclear
Physics, Erice, 1975, edited by A. Zichichi ( Plenum, New York, 1977 );
K. G. Wilson, Phys. Rev. D 10 (1974) 2445.

\bibitem{no-go}  H.B. Nielsen and N. Ninomiya, Nucl. Phys. B185, 20 (1981);
B193, 173 (1981).

\bibitem{twc98:9a} T.W. Chiu, Phys. Lett. B445 (1999) 371.

\bibitem{gwr} P. Ginsparg and K. Wilson, Phys. Rev. D 25 (1982) 2649.

\bibitem{hn97:7} H. Neuberger, Phys. Lett. B417 (1998) 141;
Phys. Lett. B427 (1998) 353.

\bibitem{ph98:2} P. Hasenfratz, Nucl. Phys. B525 (1998) 401.

\bibitem{chand98:5} S. Chandrasekharan, "Lattice QCD with Ginsparg-Wilson
fermion", hep-lat/9805015.

\bibitem{twc98:10a} T.W. Chiu, "Topological phases in Neuberger-Dirac
operator", hep-lat/9810002.

\bibitem{twc99:1} T.W. Chiu and T.H. Hsieh, "Perturbation calculation
of the axial anomaly of Ginsparg-Wilson fermion", hep-lat/9901011.

\bibitem{ml98:8b} M. L\"uscher, Nucl. Phys. B538 (1999) 515.

\bibitem{kiku98:6} Y. Kikukawa and A. Yamada, Phys. Lett. B 448, 265 (1999).

\bibitem{fuji98:11} K. Fujikawa, Nucl. Phys. B546, 480 (1999).

\bibitem{adams98:12} D. Adams, "Axial anomaly and topological charge in
lattice gauge theory with overlap-Dirac operator", hep-lat/9812003.

\bibitem{suzu98:12} H. Suzuki, "Simple evaluation of chiral Jacobian
with overlap-Dirac operator", hep-th/9812019.

\bibitem{ph98:1} P. Hasenfratz, V. Laliena and F. Niedermayer,
                 Phys. Lett. B427 (1998) 125.

\bibitem{ml98:2} M. L\"uscher, Phys. Lett B 428 (1998) 342.

\bibitem{theta} C. Callan, R. Dashen and D. Gross, Phys. Lett. 63B, 334 (1976);
                R. Jackiw and C. Rebbi, Phys. Rev. Lett. 37, 172 (1976).

\bibitem{rn95} R. Narayanan and H. Neuberger, Nucl. Phys. B443, 305 (1995).

\bibitem{kaplan92} D.B. Kaplan, Phys. Lett. B288, 342 (1992).

\bibitem{slavnov93} S.A. Frolov and A.A. Slavnov,
                    Phys. Lett. B309, 344 (1993).

\bibitem{twc98:4} T.W. Chiu, Phys. Rev. D58 (1998) 074511.

\bibitem{karsten81} L.H. Karsten and J. Smit, Nucl. Phys. B183, 103 (1981).

\bibitem{kerler81} W. Kerler, Phys. Rev. D 23, 2384 (1981).

\bibitem{seiler82} E. Seiler and I.O. Stamatescu, Phys. Rev. D25, 2177 (1982);
erratum: D26, 534 (1982).

\bibitem{fuji84} K. Fujikawa, Z. Phys. C25, 179 (1984).

\bibitem{sachs_wipf} I. Sachs and A. Wipf, Helv. Phys. Acta 65, 652 (1992).

\bibitem{hn98:3} H. Neuberger, Phys. Rev. D 59, 085006 (1999).

\bibitem{fuji79} K. Fujikawa, Phys. Rev. Lett. 42, 1195 (1979).


\end{thebibliography}
\end{document}